\documentclass[reprint,unsortedaddress]{revtex4-1}

\usepackage{graphicx,graphics}
\usepackage{amssymb,amsfonts,amsmath}
\usepackage{color}

\usepackage{graphicx}
\usepackage{amssymb}
\usepackage{amsmath}
\usepackage{epsfig}
\usepackage{chemarr}
\usepackage{url}
\usepackage{natbib}
\usepackage{dcolumn}
\usepackage{bm}

\def\beq{\begin{equation}}
\def\eeq{\end{equation}}

\def\beqa{\begin{eqnarray}}
\def\eeqa{\end{eqnarray}}
\def\gexpg{\langle \beta \Delta e^{-\beta \Delta}\rangle}

\def\ngexpg{\langle \beta \Delta e^{-\beta \Delta +1}\rangle}
\def\expg{\langle e^{-\beta \Delta}\rangle}

\begin{document}
\title{The Information Capacity of Specific Interactions}
\author{Miriam H. Huntley$^{1,*}$, Arvind Murugan$^{1,2,*}$ and Michael P. Brenner$^{1,3}$}
\affiliation{ $^1$Harvard John A. Paulson School of Engineering and Applied Sciences and the Kavli Institute for Bionano Science and Technology,  Harvard University, Cambridge, Massachusetts 02138, USA}
\affiliation{${}^2$ Physics and the James Franck Institute, University of 
Chicago, Chicago, IL 60637} 
\affiliation{$^3$Department of Physics, Harvard University, 17 Oxford St. Cambridge,  Massachusetts 02138, USA}
\thanks{These authors contributed equally to this work.}
\begin{abstract} 
Specific interactions are a hallmark feature of self-assembly and signal-processing systems in both synthetic and biological settings. Specificity between components may arise from a wide variety of physical and chemical mechanisms in diverse contexts, from DNA hybridization to shape-sensitive depletion interactions. Despite this diversity, all systems that rely on interaction specificity operate under the constraint that increasing the number of distinct components inevitably increases off-target binding.
Here we introduce `capacity', the maximal information encodable using specific interactions, to compare specificity across diverse experimental systems, and to compute how specificity changes with physical parameters. Using this framework, we find that `shape'-coding of interactions has higher capacity than chemical (`color') coding because the strength of off-target binding is strongly sublinear in binding site size for shapes while being linear for colors. We also find that different specificity mechanisms, such as shape and color, can be combined in a synergistic manner, giving a capacity greater than the sum of the parts.
\end{abstract} 

\maketitle

\textbf{Significance Statement}
The past fifteen years have seen a proliferation of experimental techniques aimed at engineering self-assembled structures. These bottom-up techniques rely on specific interactions between components that arise from diverse physical mechanisms such as chemical affinities and shape complementarity attraction. Due to this diversity of techniques, comparison of specificity across systems, each with unique physics and constraints, can be difficult. Here we describe a general quantitative framework based on information theory to compare specificity in a range of recent experimental systems. We calculate the capacity, the maximal amount of information that can be encoded using a system of specific interactions and still be resolved by the interactions, as a function of experimentally tunable parameters. Our framework can be applied to a diverse range of systems with specific interactions, from novel colloidal experiments to protein interaction data.

\section{Introduction}
Specific interactions between many species of components is the bedrock of biochemical function, allowing signal transduction along complex parallel pathways and self-assembly of multi-component molecular machines. Inspired by their role in biology, engineered specific interactions have opened up tremendous opportunities in materials synthesis, achieving new morphologies of self-assembled structures with varied and designed functionality. The two major design approaches for programming specific interactions either use chemical specificity or shape complementarity. 
 
{\sl Chemical specificity} is achieved by dividing binding sites into smaller regions, each of which can be given one of $A$ ``colors'', or unique chemical identities. Sites bind to each other based on the sum of the interactions between corresponding regions. For example, a recent two color system paints the flat surfaces of 3-dimensional polyhedra with hydrophobic and hydrophilic patterns \cite{randhawa2010importance}, or with a pattern of solder dots \cite{gracias2000forming}, allowing polyhedra to stick to each other based on the registry between their surface patterns. Another popular approach uses DNA hybridization, where specific matching of complementary sequences has been used to self-assemble structures purely from DNA strands \cite{Winfree:1998vl,Ke:2012jd}, and from nanoparticles coated with carefully chosen DNA strands \cite{Mirkin:1996th,Alivisatos:1996vs, Valignat2005,Biancaniello:2005p5460,Nykypanchuk:2008cp}.
  
{\sl Shape complementarity} uses the shapes of the component surfaces to  achieve specific binding, even though the adhesion is via a nonspecific,  typically short-range potential. In the synthetic context, shape-based  modulation of attractive forces over a large dynamic range was first proposed  and experimentally demonstrated for colloidal particles  \cite{Mason:2002cp,mason2010process} using tunable depletion forces \cite{zhaomason-roughness:2007, zhaomason-roughnessComp:2008}. Recent  experiments have explored the range of possibilities opened up by such ideas,  from lithographically designed planar particles  \cite{mason}  with undulating  profile patterns to ``Pac-man'' particles with cavities that exactly match  smaller complementary particles \cite{pine}.  The number of  possible shapes that can be made using these types of methods depends on fabrication  constraints but the possibilities can be quite rich  \cite{Murray-2010,oildroplets}. Using only non-specific surface attraction,  experiments have achieved numerous and complex morphologies such as clusters,  crystals, glasses, and superlattices  \cite{ZhaoMasonCrystals,  Rossi28042015,Mason:2002cp,Liberato-2011,G-Ra_2008}.	
   
A further class of programmable specific interactions combines both chemical specificity and shape complementarity. The canonical example is protein binding interactions \cite{Johnson:2011kh}; the binding interactions between two cognate proteins are specified by their amino acid sequence, which programs binding pockets with complex shape and chemical specificity. Recent efforts \cite{King:2012gc,Lai:2012vh} aim to rationally design these protein interactions for self-assembly. Since both the shape of the binding pocket and its chemical specificity is determined by the same amino acid sequence, these two features cannot be controlled independently. Other synthetic systems offer the promise of independent control of chemical and shape binding specificity, giving a larger set of possible interactions.

These diverse systems achieve specific interactions through disparate physical mechanisms, with different control parameters for tuning binding specificity. However, they must all solve a common problem\cite{Perelson:1979ty,Segel:1989th}: create a family of $N$ ``lock'' and ``key'' pairs that bind well within pairs but avoid off-target binding across pairs (`crosstalk'). Any crosstalk limits the efficacy of the locks and keys. For example, in the context of DNA-based affinities, although there are $4^L$ unique sequences of length $L$, the strong off-target binding severely restricts the number that can be productively used. Analogously, for colloidal systems driven by depletion interactions, there can be significant off-target binding due to partial contact. The performance of a system of specific interactions depends acutely on how the system constraints (e.g. number of available bases, fabrication length scale, etc.) limit its ability to avoid crosstalk.

In this paper, we develop a general information theory-based framework for quantitatively analyzing specificity in both natural and synthetic systems. We use a metric based on mutual information to derive a bound on the number of different interacting particles that a system can support before crosstalk overwhelms interaction specificity. Increasing the number of nominally distinct pairs beyond this limit cannot increase the effective number of distinguishable species. We compute this information-theoretic `capacity' for different experimental systems of recent interest, including DNA-based affinities and colloidal experiments in shape complementary. We show that shape-based coding fundamentally results in lower crosstalk and higher capacity than color-based coding. We also find that shape and color-based coding can be combined synergistically, giving a super-additive capacity that is greater than the sum of the color and shape parts.

\begin{figure}
\centering
\includegraphics[width=0.5\textwidth]{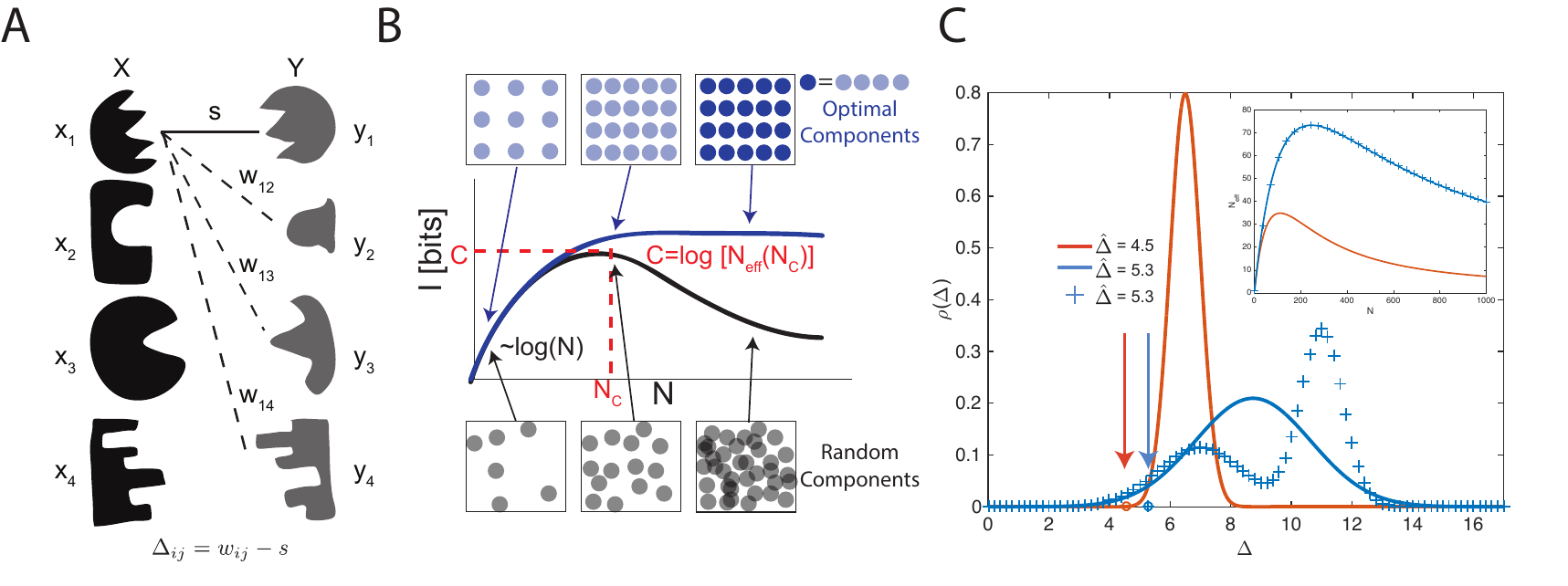}
\caption{\textbf{Information theory determines the capacity of systems of specific interactions}. \textbf{A)} A model system of locks (black) which each bind with energy $s$ to their specific key (gray) via some specific interaction. \textbf{B)} As the number $N$ of lock-key pairs is increased, non-cognate locks and keys inevitably start resembling each other as they fill up the finite space of all possible components (square boxes), both with optimized or random design of lock-key pairs. Consequently, mutual information $I$ between bound locks and keys rises with $N$ for small $N$ but reaches a point of diminishing returns at $N= N_C$; due to the rapid rise in off-target binding energy, $I$ can no longer increase, and for randomly chosen pairs, will typically decrease. The largest achievable value of $I$ is the `capacity' $C$.
\textbf{C)} Capacity $C$ can be estimated from the distribution $\rho(\Delta)$ of the gap $\Delta =  w -s$ between off-target $w$ and on-target binding energy $s$ for randomly generated lock-key pairs. 
Among the three distinct $\rho(\Delta)$ shown, the blue distributions have the same $\beta\hat{\Delta}= -\log \gexpg_\rho$. Inset: $I(N)$ (c.f., Eqn. \ref{eqn:Ihomog}), is the same for the blue distributions which, despite being markedly different in shape, have the same $\hat{\Delta}$, which captures the essential aspects of crosstalk. 
\label{fig:Fig1_GUF}}
\end{figure}

\section{The Capacity of Random Ensembles}
We consider systems where every component is designed to interact specifically with a single cognate partner, while interactions between ``off-target'' components are undesirable crosstalk. We assume that $N$ distinct ``locks'' $x_1$, $x_2$,..., $x_N \in X$, have unique binding partners, ``keys'' $y_1$, $y_2$,..., $y_N \in Y$ (Fig.~\ref{fig:Fig1_GUF}A). 
The physics of a particular system determines the binding energy $E_{ij} \equiv E(x_i,y_j)$ between every lock and key.
Assuming equal concentrations of locks and keys in a well-mixed solution, binding between lock $x_i$ and key $y_j$ will occur with probability $p(x_i, y_j) = e^{-\beta E_{ij}}/Z$ where $Z$ is a normalization factor such that $\sum_{i,j} p(x_i,y_j) = 1$ (see SI) and $\beta^{-1} = k_B T$ is the temperature scale. The mutual information $I(X;Y)$ transmitted through binding is defined as
\begin{equation}
I(X;Y)= \sum_{x_i\in X,y_j\in Y} p(x_i,y_j) \log_2 \frac{p(x_i,y_j)}{p(x_i)p(y_j)}
\label{eqn:Idef} 
\end{equation}
where $p(x_i)$ is the marginal distribution of $x_i$, representing the total probability of seeing $x_i$ in a bound pair (and similarly $p(y_j)$). 
Mutual information $I(X;Y)$ is a global measure of interaction specificity in systems with many distinct species; it quantifies how predictive the identity of a lock $x_i$ is of the identity of a key $y_j$ found bound to it. 

Consider a set of interacting lock-key pairs for which $E_{ii}=s$ for all cognate pairs (strong binding), while for crosstalking interactions (weak binding) $E_{ij}=w_{ij}=s+\Delta_{ij}$. We assume $\Delta_{ij}$ are i.i.d. random numbers drawn from a distribution of gap energies $\rho(\Delta)$, with $\Delta>0$, where the exact form of $\rho(\Delta)$ depends on the physics of the system. Denoting $\langle \rangle$ as an average with respect to $\rho(\Delta) $,
 one can approximate Eqn.~\ref{eqn:Idef} as 
 \beqa
I&=&\log_2 N_{eff}(N),\\
N_{eff}(N) &=& \frac{N}{1+(N-1)\expg}e^{-\frac{(N-1)\gexpg}{1+(N-1)\expg}} 
\label{eqn:Ihomog}
\eeqa
(see SI).
In a system with crosstalk that contains $N$ nominally distinct lock-key pairs, $N_{eff}(N)$ is the effective number of fully distinguishable lock-key pairs. 
$N_{eff}$ can be much smaller than $N$ if crosstalk is significant (e.g., if $\expg \sim O(1)$).
 
Intuitively, information theory predicts that a system with $N_{eff}(N)$ 
non-crosstalking lock-key pairs can perform a task with the same effectiveness 
as a system with $N$ crosstalking species. For example, in the self-assembly of 
a multi-component structure, distinct but crosstalking 
species can take each 
other's place, decreasing the effective number of species. This effect has been 
shown to reduce self-assembly yield 
\cite{murugan2015undesired,Jacobs:2015bs,Hedges:2014cs}. Similarly, the 
efficacy of $N$ parallel signaling pathways is known to be reduced by crosstalk 
\cite{LaubSpecificityReview}.
In Fig.~\ref{fig:Fig1_GUF}B we show a typical plot of $I = \log_2 N_{eff}(N)$. $N_{eff}$ grows initially with $N$, but stops growing at $N \sim N_C$, the point of diminishing returns; adding any further species beyond $N_C$ only increases the superficial diversity of species but cannot increase $N_{eff}$. 

Paralleling Shannon's theory of communication, we define `capacity' $C$ as
\beq
C \equiv \max_N I = \log_2 N_{eff}(N_C)
\eeq
(Fig.~\ref{fig:Fig1_GUF}B)\footnote{`Capacity' in information theory is often measured in bits/second, whereas here we intentionally use the same units as $I$. Furthermore, capacity is traditionally defined as a maximum over all possible distributions $p(X)$; here we restrict to maximizing only over one parameter, $N$, where all $N$ pairs are randomly chosen from the ensemble (see SI).}. The capacity is the largest number of bits of information that can be encoded using a system of specific interactions and still be uniquely resolved by the physics of interactions. Determining $C$, or equivalently the largest value of $N_{eff}$, is of crucial importance to both synthetic and biological systems since it limits, for example, the number of independent signaling pathways or the complexity of self-assembled structures. 

 We can compute capacity for any crosstalk energy distribution $\rho(\Delta)$ by finding the maximum of Eqn. \ref{eqn:Ihomog}. A useful approximation is 
\beq
C \approx  -\log_2  \ngexpg,  
\label{eqn:ClinD} 
\eeq  
giving a simple rule for the dependence of capacity on the binding energy distribution (see SI). The importance of maximizing $\beta \hat{\Delta} \equiv - \log \gexpg $ in Eqn. \ref{eqn:ClinD} is intuitive: in order to increase the capacity of the system, the (exponential average of the) gap between on-target $\beta \bar{s} \equiv -\log\langle e^{-\beta s} \rangle$ and off-target binding $\beta \bar{w} \equiv -\log\langle e^{-\beta w} \rangle$ should be made as large as possible (see SI for precise relationship between $\hat{\Delta}, \bar{s}$ and $\bar{w}$). 
Fig.~\ref{fig:Fig1_GUF}C shows three distinct probability distributions, two of which have identical $\hat{\Delta}$. As predicted, $N_{eff}$ reaches a higher maximum for distributions with larger $\hat{\Delta} $. 

We note that our definition of capacity uses equilibrium binding probabilities and hence applies only at long times compared to unbinding times. In practice, this typically limits $|s|\leq$ $10 \;k_BT$, and so we use this bound on $s$ herein. The formalism can be easily extended to include kinetic effects by computing $p(x_i,y_j)$ at a finite time $t$, though this is not our focus here. 

In what follows, we show how capacity depends on binding interactions and fabrication constraints for several systems of recent interest. In most systems, the on-target binding energy typically strengthens with the binding surface area $S$ of cognate pairs as $\bar{s}= - \epsilon S$, where $\epsilon$ is the binding energy per unit area. However, we find that the off-target energies $\bar{w}$ can grow with $S$ at very different rates across several systems we study. We parameterize this variation as
\begin{equation}
\bar{w} = - \epsilon \alpha S^\gamma
\end{equation}
where $\alpha,\gamma$ depend on the details of binding interactions. We show below that if the specificity is determined purely by `colors' (i.e., chemical identities), then $\gamma = 1 $. In contrast, if specificity arises from shape complementarity, $\gamma \approx 0$, as long as the range of the surface attraction is small compared to the length scale of shape variation. Thus crosstalk grows very slowly with the number of independent binding units in shape-based systems, allowing for a dramatic decrease in crosstalk and improvement of capacity relative to systems that use chemical specificity.

\section{The Capacity of Color}
\begin{figure}
  \centering
   \includegraphics[width=0.5\textwidth]{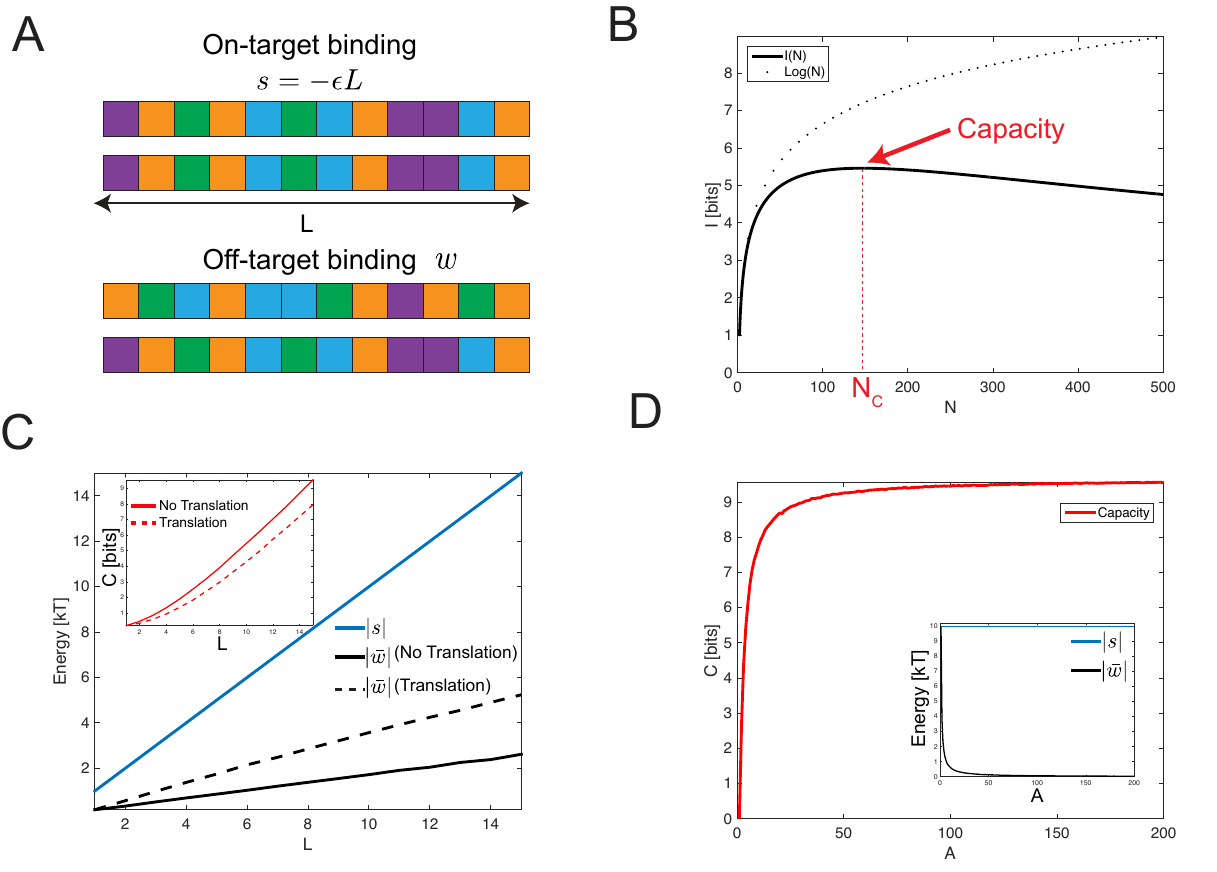}
 \caption{\textbf{Complementary color components demonstrate the capacity of programmable interactions}. \textbf{A)} Each lock $x_i$ has $L$ distinct units, each of which can be one of $A=4$ `colors' or chemical identities. Each color has a strong affinity $\epsilon$ for itself. Cognate locks and keys have the same sequence of colors and bind with energy $s= -\epsilon L$. Off-target binding energy $w$ given by the number of accidental color matches; $w = -5 \epsilon$ in the example shown. \textbf{B)} Mutual information as a function of $N$, the number of lock-key pairs. $I$ increases initially as $\log(N)$, but then reaches a maximum value, the capacity, and then decreases. ($L=10$, $\epsilon=1\; k_B T $, $A=4$). \textbf{C)} Increasing $L$ increases both the on-target strength $|s|$ and the off-target strength $|w|$. Inset: Capacity scales linearly with $L$. If translation is allowed, $|s|$ is unaffected, but $|w|$ is higher (black, dashed), and therefore the capacity is lower (red, dashed). ($L=10$, $\epsilon=1 \;k_B T$, $A=4$). \textbf{D)} Increasing the alphabet size $A$ does not affect on-target binding $s$, but does decrease $|w|$, thereby increasing $\hat{\Delta}$. ($L=10$, $\epsilon=1 \;k_B T$) \label{fig:DNA}}
\end{figure}

We first consider the capacity of interactions mediated through binding sites which are subdivided into multiple regions, each of which can be assigned any one of $A$ chemical identities or ``colors''. We take inspiration from DNA coding that acts via complementary hybridization between single stranded DNA. Previous work \cite{Milenkovic} developed engineering principles for determining the optimal length and nucleotide composition of these DNA strands based on detailed models of the binding energy. Information theoretic measures have also been used to understand binding of transcription factors to DNA and other sequence-based molecular recognition problems \cite{Myers:2008gi,Schneider:1997, itzkovitz2006coding, Savir:2009ul, WuChaikin:2012}. While the theory of DNA coding has a long history \cite{adleman1994molecular}, our contribution here formulates the problem in a mutual information framework that relates the capacity to a physical quantity and hence allows for direct comparison of varied chemical (`color') and shape systems. 

In our simplified color model, a lock is composed of $L$ units, each of which is painted with one of $A$ chemical colors (Fig. ~\ref{fig:DNA}A). Each color binds to itself with energy $-\epsilon $ and binds to other colors with energy $0$, such that locks and their cognate keys have the same sequence. The binding energy of any two strands $x_i$ and $y_j$ is given by $E_{ij} = \sum_{l=1}^L -\epsilon \delta_{x_i^l,y_j^l}$ (where $x_i^l$ is the color of the $l$th site of $x_i$, and $\delta$ is the Kronecker delta). 
We analyze this system with translations, where $E_{ij}$ is given by the strongest binding across all possible translations of the two strands relative to each other, as well as without translations.

We calculate $I(N)$ by sampling $N$ randomly selected pairs of locks and keys, constructing the interaction matrix $E$, and computing $I(N)$ using Eqn. \ref{eqn:Idef}. We average $I(N)$ over many repetitions. 
An approximate but faster method to compute $I(N)$ (necessary for large $L,N$) uses Eqn. \ref{eqn:Ihomog}, sampling random pairs of off-target locks and keys to estimate $\expg$ and $\gexpg$. The two methods give nearly identical results (see SI), and the calculations in the paper henceforth are carried out with the second method. 

Fig.~\ref{fig:DNA}B examines $I(N)$ when $L=10$, $A=4$, and $\epsilon=1 \; k_B T$ so that a lock and its key bind together with on-target energy $\beta s = -10$. The mutual information has a maximum of 5.5 bits near $N_C=146$, far less than the total number of unique sequences ($4^{10} = 1,048,576$). Due to crosstalk, even though there are nominally $146$ pairs at capacity, the system behaves as if there are only $N_{eff}=44$ independent pairs.

An obvious way of increasing capacity is to boost $\hat{\Delta}$ by increasing $L$. This strengthens both on-target binding and off-target binding, since both $s$ and $w$ scale with $L$ ($\beta\bar{w} = -L \log\frac{A-1+e^{\beta \epsilon}}{A}$). However, the gap between them widens, and the capacity scales linearly with $L$ (Fig.~\ref{fig:DNA}C solid line) (see SI). As a comparison, we also show the capacity when translation is allowed between any two strands. 
Off-target strands can now translate until they find the strongest binding, increasing crosstalk and thus lowering capacity.

In practice, on-target binding $|s|$ must be limited to below approximately 10 $k_BT$ for the binding to be reversible; hence $L$ cannot be increased arbitrarily without also decreasing $\epsilon$. An alternate way to increase capacity at fixed $s$ is to increase the number of `colors' $A$. As $A\rightarrow \infty$, accidental mismatches in off-target binding are rare; $|w| \rightarrow 0$, and the capacity is only limited by $s$. In Fig.~\ref{fig:DNA}D, capacity in the large $A$ limit can be approximated by setting $\beta \Delta=-s=10$ in Eqn.~\ref{eqn:ClinD}, giving $C=9.6$ bits. However, in practice, alphabet size $A$ cannot be easily increased in experiments, and other techniques must be used to decrease the off-target binding strength, such as the use of shape complementarity. 

\section{The Capacity of Shape}	
\begin{figure}
 \centering
  \includegraphics[width=0.5\textwidth]{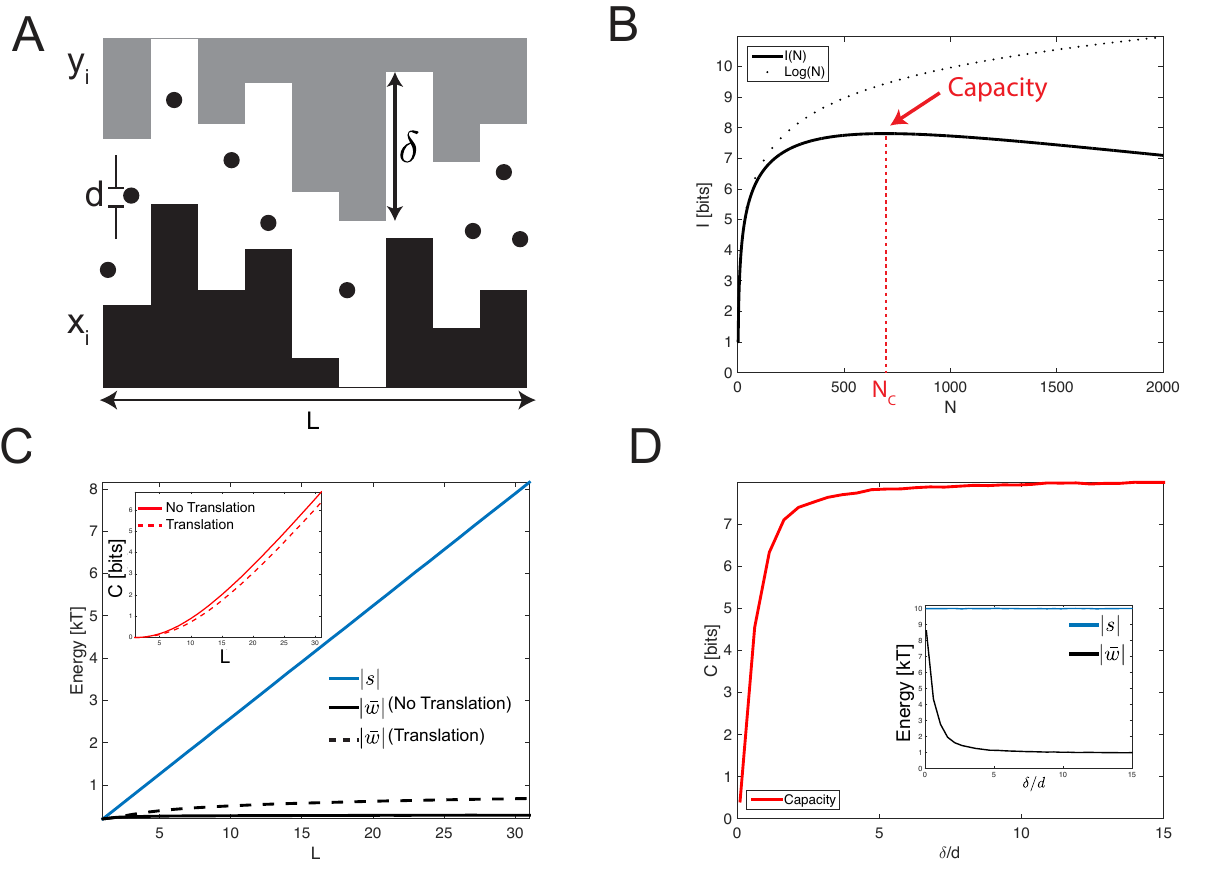}
 \caption{\textbf{Complementary shapes demonstrate the capacity of programmable interactions}. \textbf{A)} Each shape $x_i$ is made of $L$ vertical bars of different heights and has a corresponding binding partner $y_i$ shaped exactly as its complement; interactions are mediated via depletant particles of diameter $d$, and each adjoining bar can change by a maximum amount of $\delta$. \textbf{B)} Mutual information as a function of $N$, the number of lock-key pairs, showing a capacity of 7.8 bits. ($L=10$, $d=0.2\;\mu$m, $\delta = 1\;\mu$m, $
 \bar{ s } = -10\; k_B T$). 
 \textbf{C)} Increasing $L$ increases the on-target binding strength $|s|$ (blue), but has little effect on off-target binding strength $|w|$ (black), unlike with colors (Fig.~\ref{fig:DNA}C). Inset: Capacity scales linearly with $L$. Allowing translations has no effect on $|s|$, but increases $|w|$ (black, dashed) and therefore decreases $C$ (red, dashed). ($d=0.07\;\mu$m, $\delta = 1\;\mu$m, $\epsilon=1 \; k_B T$). \textbf{D)} Fixing $\bar{s}=-10\; k_B T$, the capacity can be increased by decreasing $\delta$: when $\delta/d$ is small, on-target keys are indistinguishable from off-target keys, and so capacity is small. Increasing $\delta$ decreases the crosstalk, and capacity increases accordingly.\label{fig:Tetris} }
\end{figure}

Systems of interacting, complementary shapes are characterized by the non-specific binding of surfaces mediated by a short-range force of characteristic length $\lambda_{\text{shape}}$. The components' shapes sterically allow or inhibit two surfaces from coming into contact, dictating specificity. We find that crosstalk is qualitatively weaker in such shape-based systems, resulting in higher capacity than color-based models. 

We examine the capacity of a model inspired by a recent experimental system consisting of lithographically sculpted micron-sized particles with complementary shapes \cite{mason} whose attractive interactions are mediated by the depletion force. The constraints on the shapes of these components (size $< 10 $ $\mu$m, line width $> 400$ nm, radius of curvature $> 200$ nm) still leave a large variety of shapes that can interact in a lock and key fashion, yet crosstalk between similarly shaped components reduces the number of effectively unique pairs. We model this system by defining each solo component as a series of $L$  adjoining bars of various heights, whose profile is similar to a Tetris piece.  For each lock $x_i$, the shape of the cognate key $y_i$ is exactly  complementary, as in Fig.~\ref{fig:Tetris}A. We account for fabrication  constraints by setting the width of each bar to $1$ $\mu$m and restricting the  change of one bar height relative to its neighboring bars to be less than  $\delta = 1 $ $\mu$m. Depletant particles of diameter $d$ (typically $100-200$  nm) create an attractive energy of $-\epsilon (d-h)$ for two surfaces  separated by $h < d$. Thus $\lambda_{\text{shape}} \sim d$. In experiments,  $\epsilon$ is set by the depletant particle volume fraction and the  temperature. In principle, the fabrication fidelity must also be accounted  for, as local defects in the shape will disrupt cognate binding. The effect of such defects is shown in the SI; we find that defects of size much less than $d$,  the depletant particle size, have minimal impact on capacity. We assume such a limit in the remainder of the text. 

We find that crosstalk with shapes differs fundamentally from the color models discussed earlier. While on-target binding strength still increases linearly with $L$, off-target binding is almost independent of $L$ (Fig.~\ref{fig:Tetris}C). In fact, we find that for large enough L, off-target binding $\bar{w} \sim -L^0$; for larger $\delta/d$ (or smaller $L$), $\bar{w}$ is still strongly sublinear in $L$ (see SI). The weak dependence of $\bar{w}$ on $L$ can be understood intuitively, as a lock pressed to a random mismatched key will typically come into contact at a single location. In contrast, in color-based systems, off-target locks and keys are in full contact and hence $\bar{w} \sim -L$. Thus $\hat{\Delta}$ and hence capacity $C$ for shape systems can be significantly higher than for color based systems with the same strong binding energy $\bar{s}$. In Fig.~\ref{fig:Tetris}B, $C_{shape}=7.8$ bits while $C_{color}=5.5$ bits with similar parameters ($\bar{s}=-10 \; k_B T$, $L=10$). Finally, in Fig.~\ref{fig:Tetris}D, for fixed $L$, we find that capacity falls rapidly and all specificity is lost when the spatial range of depletion interactions $\lambda_{\text{shape}} \sim d$ exceeds the scale of spatial features $\delta$, as expected. These results are consistent with earlier experiments \cite{zhaomason-roughness:2007} and computational models \cite{zhaomason-roughnessComp:2008} that established a high dynamic range in the strength of depletion interactions between surfaces roughened by asperities, and in particular found that the attraction between surfaces was diminished when the asperity height was below the depletion particle size.

Our results, while intuitive in retrospect, point to a qualitative advantage for coding through shapes; random mismatched shapes have a crosstalk that is, at worst, sublinear in binding site size while crosstalk is linear in site size for color-based systems. Our work suggests that such increased specificity is very robust as it is derived from basic properties of shape itself. Knowing the precise benefits of shape-based coding is important in deciding to incorporate it in engineering efforts going forward.

\subsection{Lock and Key Colloids}
We may further apply this framework to the recent experimental system of lock-key colloids. In this system \cite{pine}, a key is a sphere of radius $r$ (typically ~ 1-3 microns), while its cognate lock is a larger sphere with a hemispherical cavity of radius $r$, complementary to its key (see SI). The attraction is mediated by depletant particles of diameter $d\approx 50-100$ nm. Multiple pairs of locks and keys may be used concurrently, with the $i$th pair having a key radius of $r_i$, with the risk of keys binding to incorrect locks.

\begin{figure}
 \centering
\includegraphics[width=0.5\textwidth]{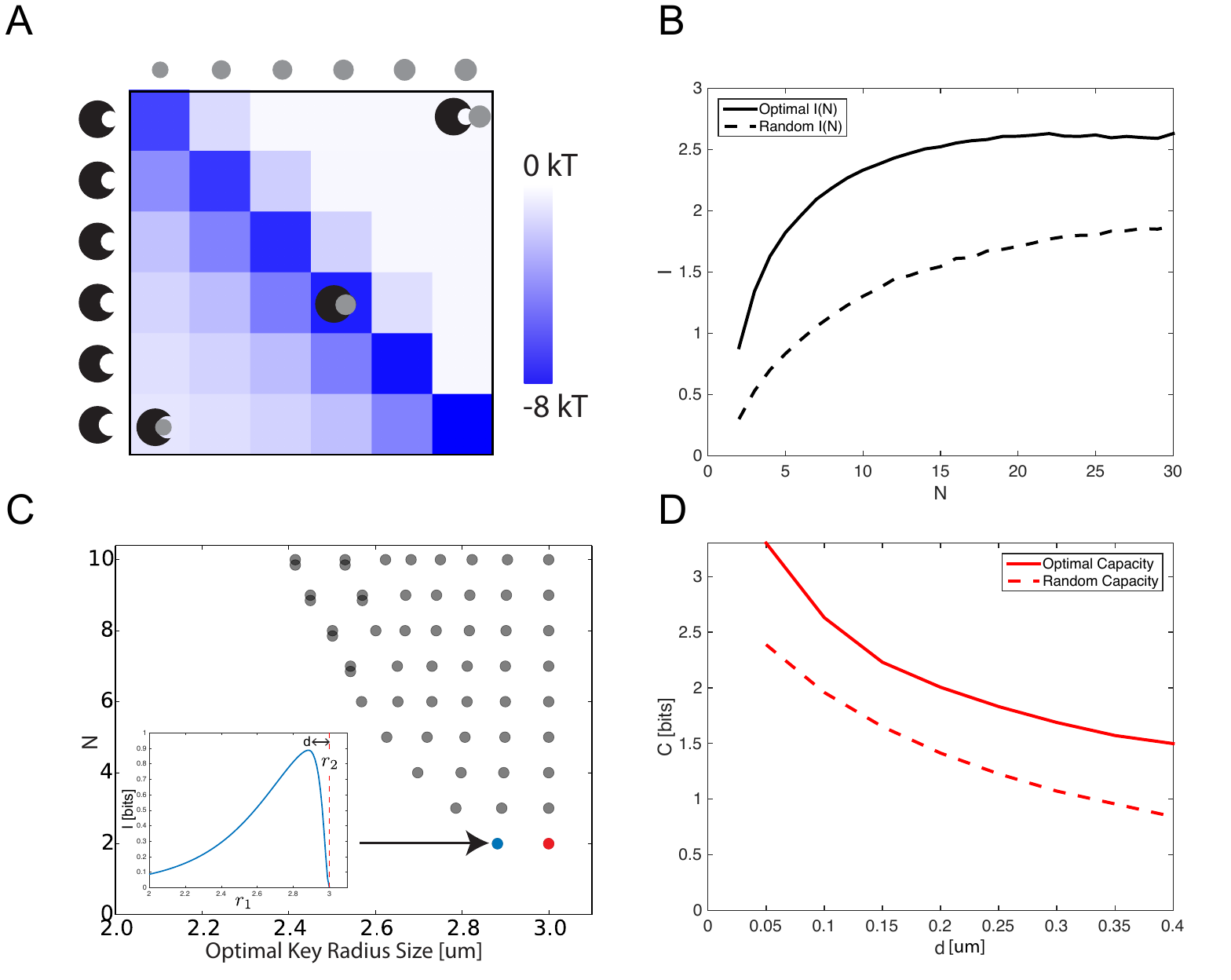}
 \caption{\textbf{Pacman lock/key pairs demonstrate the capacity of shape space}. \textbf{A)} Interaction energies of 6 optimally selected lock-key pairs. Cognate locks and keys fit snugly, while crosstalk is most severe between small keys and larger locks. \textbf{B)} Mutual information plotted as a function of the number of pairs $N$ shows that both optimal (solid) and random (dashed) sets of $\{r_i\}_N$ display a maximum in mutual information. (Random pairs are drawn uniformly from [1,3] $\mu$m.) \textbf{C)} Each row, plotted at $y=N$, shows the optimal $\{r_i\}_N$  for $N$ lock-key pairs. After saturation ($N>6$), particles are duplicated (overlapping circles). Inset: Mutual information with $r_2$ held fixed at $3$ $\mu$m shows how mutual information varies as $r_1$ changes. ($d=100$ nm, $s_{\text{max}} = -8 \;  k_B T$). \textbf{D)} The capacity increases with smaller depletant particle size. On-target binding at $r=R_{max}$ is fixed to $-8 \;  k_B T$.\label{fig:Fig2_Pine}}
\end{figure}

How should one choose $N$ lock-key radii $r_i$ to minimize crosstalk and maximize capacity?
We may gain some intuition by considering a system containing only two lock-key pairs of radius $r_1$ and $r_2$ respectively. The on-target binding energies of the two pairs are proportional to the area of contact: $E_{11} \sim r_1^2$, $E_{22} \sim r_2^2$ since each key makes perfect contact with its own lock. Assuming $r_1 < r_2$, crosstalk $E_{12} \sim r_1$, corresponding to the larger key of size $r_2$ contacting an annulus around the smaller lock of size $r_1$. The other crosstalk energy $E_{21} \sim r_2^2 - (r_2 - r_1) f(r_2,d)$ is typically much larger, corresponding to the smaller key fitting into the larger lock of size $r_2$ (see SI for complete derivation). Thus, there are two competing pressures on the radii $r_1,r_2$: increasing the overall size of both pairs $r_1, r_2$ improves specificity since the on-target energies $r_1^2,r_2^2$ grow faster than the crosstalk terms. Yet $E_{21}$ grows rapidly if the radii are too similar to each other. Hence the optimal solution for $N=2$ requires setting $r_2 = R_{max}$ (the largest allowed radius) and $r_2 -r_1 \approx d$. The binding energy of 6 particles (in this case optimally chosen to maximize $I$) is shown in Fig~\ref{fig:Fig2_Pine}A, with on-target binding and the two types of off-target binding shown.

This intuitive argument does not capture many-body effects that determine capacity for larger $N$. We find the optimal $\{r_i\}_N$ at fixed $N$ by maximizing the mutual information $I$ in Eqn. \ref{eqn:Idef} numerically through gradient descent; note that Eqns. \ref{eqn:Ihomog} cannot be used since the on-target binding energy $s$ varies across pairs.
Fig.~\ref{fig:Fig2_Pine}B (solid line) shows the mutual information of optimally chosen radii as a function of $N$, an improvement over randomly chosen radii (dashed line). Fig \ref{fig:Fig2_Pine}C shows the optimal set of radii for various $N$, with $d=100$ nm and $R_{max}=3\; \mu m$; the optimal spacing of the radii is $O(d)$.

Interestingly, when $N > 6$, the system has exceeded its capacity. $I$ does not increase any further (Fig \ref{fig:Fig2_Pine}B) and the optimal set involves repeating locks and keys of the smallest radii. Intuitively, the smallest lock-key pairs have become so small that making an additional lock-key pair of an even smaller radius would yield very low self-binding energy relative to the incurred crosstalk. Hence the only way to increase $N$ without decreasing $I$ is to create new nominal pairs at the smallest radius; such pairs are obviously indistinguishable through physical interactions and hence do not increase mutual information any further. We find that this capacity decreases with increasing size of depletant particles and falls to $N_C \sim 1$ by $d = 400$ nm. Similarly, increasing $R_{max}$ (with fixed largest cognate binding energy $s = -8 \;  k_B T$) increases capacity.

Thus we find that this colloidal particle system can support about $N_C \sim 6 - 8$ lock-key pairs without much crosstalk, with depletion particles of diameter $100$ nm and restricting the largest binding energy to $s_{\text{max}}=-10 \;  k_B T$. This is far smaller than the capacity of either DNA sequences or general shape-based strategies. However, these lock-key colloidal pairs are characterized by only one parameter (the radius), so the space of  available pairs is significantly smaller than DNA or shape systems with $L$ parameters. In particular, in the current system, additional lock-key pairs are forced to be of smaller radii and hence of lower and lower cognate binding energies. Such considerations emphasize the importance of quantitative information-theoretic optimization in systems with such a limited shape space.

\section{Combining Channels}
Thus far we have focused on locks and keys interacting exclusively through a single kind of physical interaction. Using our quantitative framework, we may ask how capacity increases when multiple sources of specificity, such as shape and color, are combined in a single set of locks and keys. As is known in information theory \cite{alon1998,cover2012elements}, 
the combined capacity of two interacting channels can be significantly higher than sum of the individual capacities.

\subsection{Linking Two Systems}
The simplest model for combining two channels is to physically link a lock of system 1 to a lock of system 2. We assume that there is no interaction between the two parts of the lock, or between the key from one system with the lock of the other system. (We do not take into account entropic effects due to avidity.) Thus for a linked system (which we write as $\text{System}_1\oplus\text{System}_2$), the two independent systems with gaps of $\Delta_1 $ and $\Delta_2$ are combined such that $\Delta_{Tot} = \Delta_1+\Delta_2$. Hence the gap distribution of the linked system is the convolution of the independent systems: $\rho_{Tot}(\Delta)=\rho_1(\Delta) \ast \rho_2(\Delta)$, and the capacity can be computed using Eqn. \ref{eqn:Ihomog} in terms of the gap distributions of the individual systems. 

When two channels are linked in this form without any interaction, we expect the total capacity of the system to be $C_{Tot}=C_1+C_2$ \cite{cover2012elements}. We explicitly compute this linked capacity for the physical system shown in Fig.~\ref{fig:ShapeandColor}A (left), in which a color system of length $L$ is linked to a shape system of length $L$ (Shape $\oplus$ Color). The distribution $\rho_{Tot}(\Delta)$, obtained by convolving $\rho_{color}(\Delta)$ and $\rho_{shape}(\Delta)$ is shown in Fig.~\ref{fig:ShapeandColor}B. The resulting capacity $C_{Tot} \approx C_1+C_2$ is additive up to $\log L$ corrections that are small when $L$ is large (see SI).

\subsection{Mixing Two Systems}
\begin{figure}
 \centering
 \includegraphics[width=0.5\textwidth]{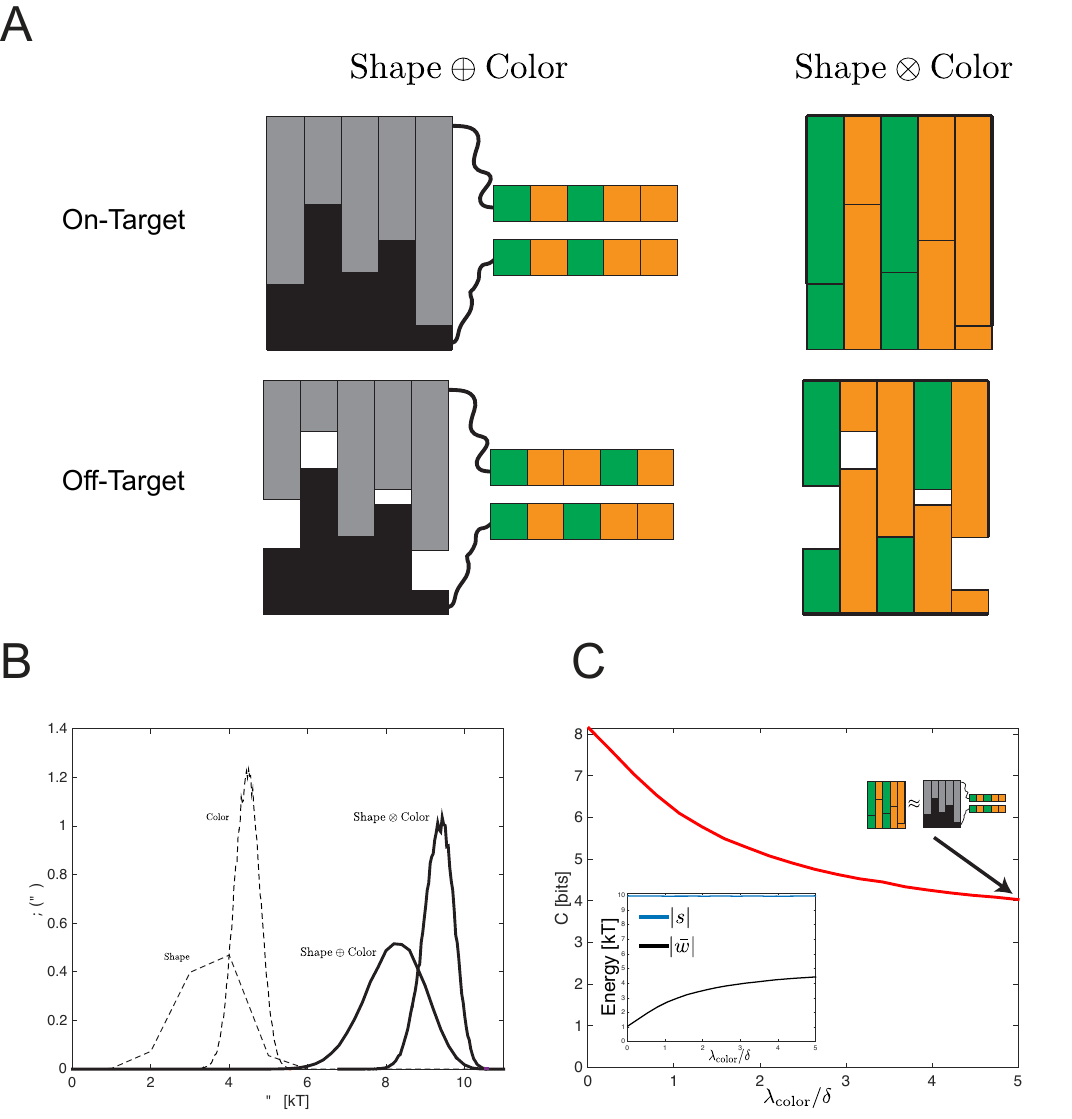}  
 \caption{\textbf{Combining Color and Shape}. \textbf{A)} Shape and color can be `linked' in an independent manner (left, $\oplus$), or `mixed' in a dependent manner (right, $\otimes$), where the shapes are coated with the chemical binding agent.
 Crosstalk in $\oplus$ results from accidental matches in \textit{either} channel, while in the $\otimes$ model, accidental matches in color contribute to crosstalk only if shapes are also matched at the same sites. \textbf{B)} As a result, the gap energy for Shape $\otimes$ Color is higher than for $\oplus$.
Here $L=10$ for both color and shape, $d=0.05$ $\mu$m, $\delta=1$ $\mu$m, $A=4$, and $\lambda_{\text{color}}=0.01$ $\mu$m. \textbf{C)} The capacity as a function of the spatial range $\lambda_{\text{color}}$ of color-based interactions. When $\lambda_{\text{color}}$ is smaller than $\delta$, the typical size of shape-based features, shape helps reduce crosstalk in color. This synergy is lost when $\lambda_{\text{color}} \sim \delta$, and color and shape act independently. ($\epsilon_{x^l = y^l}=0.5\;k_BT, \epsilon_{x^l \neq y^l} = 0.25\;k_BT$, see SI).\label{fig:ShapeandColor}}
\end{figure}

In a mixed system, the physics of the individual systems are combined, and there is no general formula for the resulting gap distribution since $\Delta_{Tot} \neq \Delta_1 +\Delta_2$. We study a model in which shapes are coated with chemical colors, and we denote mixed systems by
 $\text{System}_1\otimes\text{System}_2$ (Fig.~\ref{fig:ShapeandColor}A, right). The energy is the sum of the shape and color interactions, but the color interaction energy implicitly depends on the shape; only when the surfaces are near each other can the color-dependent interaction matter. We assume a distance dependence of the color interaction, with length scale $\lambda_{\text{color}}$, such that the energy of interaction decays as $e^{-h/\lambda_{\text{color}}}$ for two surfaces separated at a distance $h$.

We can intuitively understand how the mixed model differs from the linked model by examining random off-target pairs, as shown in Fig.~\ref{fig:ShapeandColor}A. In the $\oplus$ model, crosstalk arises from accidental matches in \textit{either} independent channel; hence the crosstalk is simply the sum of the number of matching sites in the two channels. However, in the $\otimes$ model, crosstalk in the color channel can arise at a site only when there is an accidental match in \textit{both} color and shape channels at that site. For example, in Fig.~\ref{fig:ShapeandColor}A, all three matching color sites contribute to crosstalk in the $\oplus$ model. However, in the $\otimes$ model, these three sites are not accidentally matched in the shape channel; since the three color sites are not in contact, they do not contribute to crosstalk. As a result, off-target binding is generally weaker and the typical gap $\Delta$ higher in the $\otimes$ model, as we find in Fig.~\ref{fig:ShapeandColor}B. Thus the mixing of shape and color in this interactive manner increases the capacity.

We may further examine how the capacity changes as a function of $\lambda_{\text{color}}$, the interaction range of the color system. When $\lambda_{\text{color}}$ is small compared to $\delta$, the maximum height of local shape features, shape features can be easily distinguished by the color force and so the color and shape work in concert to increase capacity. Increasing $\lambda_{\text{color}}$ blurs the shape contours and the color interactions no longer distinguish shapes, thereby becoming less specific. Indeed, Fig \ref{fig:ShapeandColor}C shows that when $\lambda_{\text{color}}/\delta$ becomes large, the color system and the shape system act independently, and the capacity relaxes to the capacity of the linked system Shape $\oplus$ Color. 

In summary, laying out color-based codes on undulating surfaces significantly reduces the total crosstalk since color-matched sites must also be matched in shape to contribute to crosstalk. Such color-shape synergy persists so long as the spatial range of color interactions is shorter than the length scale of shape variation.

\section{Discussion}
Here we have shown that mutual information provides a general metric for specificity, bounding the number of distinct lock-key pairs that can be supported by systems of programmable specific affinities. Mutual information is well suited as a measure of specificity for many reasons. First, mutual information is a global measure of specificity, accounting for all possible interactions between $N$ species of locks and keys. Second, as a result, it provides a precise answer as to how many particle pairs can be productively used in a given system. As $N$ is increased, crosstalk necessarily increases as we crowd the space of possible components (Fig.~\ref{fig:Fig1_GUF}B,\cite{Perelson1979}) with more and more lock-key pairs. Capacity is determined by the point $N=N_C$ at which the information gain due to larger $N$ is negated by the increase in crosstalk. 

Third, we can use mutual information to quantitatively compare disparate types of programmable interactions, from DNA hybridization to depletion driven interactions. Our framework can also quantitatively predict how varying physical parameters (e.g., depletion particle size, range of interactions, elastic modulus of shapes) raises or lowers specificity. The models we discuss can be further refined in various ways, for example by allowing DNA strands to fold, examining shapes in three dimensions, or taking into account the entropic effects of  multivalency and avidity \cite{kane2010thermodynamics}.

Using such an approach, we found that (1) shape complementarity intrinsically suffers less crosstalk than `color' (i.e., chemical specificity)-based interactions and (2) multiple physical interactions, such as color-based and shape-based interactions, can be combined in a synergistic manner, giving a capacity that is greater than the sum of the parts. Such predictions are especially valuable, given the proliferation of different mechanisms for creating and combining distinct mechanisms of specificity: mutual information provides an unbiased way of comparing their efficacy to each other.  As programmable specificity continues to drive technological developments in self-assembly \cite{zeravcic2014size}, understanding how the mutual information of paired components can be built up towards creating larger, multi-component objects is a critical future direction of this work. 

While we focus on applications to colloidal systems, we note that the framework developed here can be employed to study biological systems as well. In 1890, Emil Fischer proposed the `lock and key' model as an analogy for understanding enzyme-substrate specificity \cite{fischer1894einfluss}, focusing on the physical shapes of paired interacting components; mutual information encompasses this idea and can be applicable to a large number of biological systems. In particular, our model is useful for predicting the differences between interacting proteins that use shape complementarity alone and those that combine both shape and electrostatic complementarity (e.g. Dpr-DIP vs Dscam proteins \cite{dpr-dip}), and may also be applied to a host of other biological interaction networks \cite{Johnson:2011kh} where information transmission and pair specificity play critical roles in biological function (e.g. HKRR proteins \cite{laub} and the immune system \cite{Perelson1979}). Crucially, the mutual information model provided above is flexible enough to be extended to some of the challenging physics encountered in biology. Nonequilibrium systems can be accounted for by computing time dependent probabilities of interactions instead of the equilibrium probabilities, while hypotheses for increased specificity like `induced fit' and recent variants \cite{Savir:2009ul} can be tested directly for their impact on capacity.

In this work, we have shown that mutual information is a powerful tool to describe diverse specificity models. The strength of our framework is that it is broadly applicable - it may immediately be applied to any system for which the pairwise energies of interactions are known, in both biology and in synthetic experiments. We believe that using the capacity as a measure of system specificity will provide a simple metric for analyzing, comparing, and optimizing systems of programmable interactions. 

\begin{acknowledgments}
This research was funded by the National
Science Foundation through DMR-1435964, the Harvard Materials Research Science and Engineering
Center DMR1420570 and  the Division of Mathematical Sciences DMS-
1411694.
MPB is an investigator of the Simons Foundation. We thank Elizabeth Chen, Mikhail Tikhonov, Matthew Pinson, and Lucy Colwell for helpful discussions.
 \end{acknowledgments}

\end{document}